\documentstyle[12pt]{article}

\newlength{\extraspace}
\newlength{\extraspaces}
\newcommand{\be}{\begin{equation}}
\newcommand{\ee}{\end{equation}}
\newcommand{\bea}{\begin{eqnarray}}
\newcommand{\nn}{\nonumber}
\newcommand{\eea}{\end{eqnarray}}

\newcommand{\erf}{{\rm erf}}
\newcommand{\intR}{\int\limits_{-\infty}^\infty}
\newcommand{\C}{C_\epsilon}
\newcommand{\D}{D_\epsilon}
\begin{document}

\thispagestyle{empty}

\begin{flushright}
{\sc OUTP}-96-32P\\
17th  June 1996\\
 hep-th/9606102
\end{flushright}
\vspace{.3cm}

\begin{center}
{\large\sc{  D-BRANE RECOIL AND  LOGARITHMIC OPERATORS} }\\[15mm]

{\sc Ian  I. Kogan, Nick E. Mavromatos\footnote{PPARC Advanced Fellow}
 and John  F. Wheater} \\[5mm]
{\it Department of Physics, University of Oxford,\\Theoretical Physics,\\ 1 
Keble Road,\\
       Oxford, OX1 3NP, UK} \\[15mm]

{\sc Abstract}

\begin{center}
\begin{minipage}{14cm}
We construct the pair of logarithmic operators associated with
 the recoil of a $D$-brane.  This construction establishes a connection between
a translation in time and a world-sheet rescaling. The
problem of measuring the centre of mass coordinate
of the $D$-brane is considered and the relation between
the string uncertainty principle and the logarithmic operators is discussed.

\end{minipage}
\end{center}

\end{center}

\noindent

\vfill
\newpage
\pagestyle{plain}
\setcounter{page}{1}

\renewcommand{\footnotesize}{\small}

Recently it was suggested \cite{kogmav} that the world-sheet
description of the collective coordinates of a soliton in string theory
 is given by logarithmic operators \cite{gur}. This suggestion was
 based on earlier observation made in \cite{ckt} that logarithmic
 operators may correspond to  hidden continuous symmetries, which
 in the string  soliton case turn out  to be the target space symmetries 
 related to the zero modes (collective coordinates). 
 
Soliton backgrounds in string theory have received recently 
much attention as a result of Polchinski's
discovery~\cite{dbranes} that in the Ramond-Ramond sector
superstring solitons can be simply described by
open strings on a disk with Dirichlet boundary
conditions for the collective coordinates of the soliton.
Such constructions are known as $D$(irichlet)-branes,
and they are believed to be related to
ordinary closed string backgrounds
by duality transformations~\cite{dbranes}.

An important aspect of a $D$-brane quantization 
is the incorporation of proper recoil effects
 during the scattering
of closed string states off the $D$-brane 
background  \cite{recoil,periwal,recoil1}.
It has been argued in   these papers 
  that operators describing the recoil appear
as a result of extra logarithmic divergences arising in
the open string one-loop amplitudes
describing target-space quantum corrections to
the scattering of elementary string states off the $D$-brane
(soliton). These operators must be logarithmic ones as 
 was suggested in \cite{kogmav} and the aim of this letter is to
 give a precise construction of the ``recoil'' logarithmic operators
 for a $D$-brane (actually we shall consider  the $0$-brane most of the
 time). As a consequence we will obtain a new and very interesting result 
concerning the
 connection between world-sheet scale and evolution in time
 as well as a new way to look at the string uncertainty
relation.  This might be relevant to the consistent quantization of
soliton backgrounds in string theory  \cite{emn}.

It is important to stress that the logarithmic operators
come in pairs, $C$ and $D$, and the OPE of the stress-energy tensor T with 
   logarithmic operators $C$ and $D$ 
 is non-trivial and involves mixing~\cite{gur}
\bea
&~&T(z) C(w) \sim  \frac{\Delta}{(z-w)^2}C(w) + \dots \nn \\
&~&T(z)D(w) \sim \frac{\Delta}{(z-w)^2} D(w) + \frac{1}{(z-w)^2}C(w) + \dots
\label{stress}
\eea
where $\Delta $ is the conformal dimension of the
operators and appropriate normalizaton of the
$D$ operator  has been assumed.
The two point functions are given by \cite{gur}, \cite{ckt}
\bea
&~&   <C(z) C(0) > \sim 0 \nn \\
&~&   <C(z) D(0) > \sim \frac{c}{|z|^{2\Delta}} \nn \\
&~&    <D(z)D(0)> \sim \frac{c}{|z|^{2\Delta}} \log|z|
\qquad c=const
\label{two}
\eea
In the closed string case these relations are
to be understood as being accompanied by their
anti-holomorphic counterparts, whilst in the open
string case (relevant here)
$z,w$ are real parameters defined on the boundary of the
world-sheet disk.

The above relations are the
consequence of the behaviour
of the conformal blocks of the underlying conformal field theory,
the latter being determined by the four-point functions
of the theory.
Conformal blocks in theories involving logarithmic operators
exhibit logarithmic scaling violations on the world sheet
due to logarithmic divergencies.  The absence of
double or higher logarithmic divergences
implies the vanishing
of the
$C-C$ two-point function in (\ref{two}).
This is  an important point to remember
if one tries to guess the leading divergent
behaviour of the (unknown) conformal blocks of the
theory from knowledge of the logarithmic operators.
The non-trivial mixing between $C$ and $D$ operators
is, on the other hand, a characteristic
non-trivial property of the Jordan-cell structure
of theories involving logarithmic operators.
In the string-soliton
case it is this mixing that leads to
logarithmic divergencies of the annulus amplitudes~\cite{kogmav}.
As we shall discuss later on  this mixing is
associated with
the lack of unitarity of the effective
low-energy theory, in which $D$-brane
(quantum) excitations are ignored.

In this paper we shall concentrate for simplicity
on the case of  the $0$-brane. The extension to  $p$-branes
is straightforward, except for a few subtleties 
which  we will discuss at the end of the paper.
We recall that  
 the world-sheet
boundary operator describing
the excitation of a $D$-brane is (see ~\cite{callan} and references therein)
\be
   {\cal V}_D = \int _{\partial \Sigma} y_i \partial _n X^i
+ u_i X^0 \partial_n X^i
\label{dbraneop}
\ee
where $n$ denotes the normal derivative on the boundary of the world
sheet $\partial \Sigma$,
which at tree level is assumed to have the topology of a disk of size $L$;
$X^i~,i=1,\dots 9$ denote the collective excitations of the brane
satisfying Dirichlet boundary conditions
on the world-sheet boundary while  $X^0$ is the time and satisfies 
 standard Neumann
boundary conditions,
\be
 X^i ({\rm boundary}) = 0,~i=1,\dots, 9
  \qquad \partial _n X^0 ({\rm boundary}) = 0
\label{five}
\ee
The coefficient $u_i $ in (\ref{dbraneop}) is the
  velocity of the $0$-brane (point particle), and $y_i$ is the initial
position ($X^0= 0$) of the collective coordinates 
 of the $0$-brane \footnote{for flat
Euclidean world-volume $y_i=\delta _{ij}y^j$. In this work we shall
not discuss the important (and more realistic) case of curved
world volumes, where the simple Dirichlet boundary
conditions are known to be conformally non-invariant~\cite{dbranes,li}.}.
The operator (\ref{dbraneop}) describes an `eternally
moving' (boosted) $0$-brane. Alternatively (\ref{dbraneop}) can be
 thought of as generating the action of the Poincar\'e group on the 
 $0$-brane with $y_i$ parametrizing translations  and $u_i$
parametrizing boosts. 

To describe recoil we need an operator which has non-zero matrix
elements between two different states of the $0$-brane. In an impulse
 approximation this can be achieved 
by introducing a factor of  the  Heavyside function, $\Theta (X^0)$
 into  (\ref{dbraneop}).
This describes a $0$-brane that starts moving at
time  $X^0=0$. The initial position of the
$0$-brane at $X^0=0$
is assumed to be given by the $y_i$. 
  One can then   write down
the following expression for  the `impulse' operator \cite{periwal}:
\be
{\cal V}_{imp} \equiv
\int d^2z\thinspace \partial _\alpha ([u_i X^0 ]\Theta (X^0)
     \partial _\alpha X^i) =
 \int d\tau\thinspace  u_i \left( X^0 \Theta (X^0)\right)
     \partial _n X^i)~; \qquad i=1, \dots 9.
\label{wrongrecoil}
\ee
where the coupling constant $u_i$ is  the change in
velocity of the brane,
not only by physical arguments,
but also as a result of
imposing overall conformal invariance of the
annulus and disc amplitudes~(for more details see \cite{recoil}
-\cite{recoil1}).  We note  that the $\partial _n X^i$ parts
  yield  the standard $1/z^2$
world-sheet short-distance
behaviour in (\ref{two}). The non-trivial behaviour
is encoded in the $X^0$ part, and, therefore, 
in what follows 
we shall  concentrate on
the $X^0$-dependent parts of (\ref{wrongrecoil}).

Let us hypothesise that $X^0\Theta (X^0)$  plays the role of the 
 $D$ operator.  As it stands $\Theta (X^0)$ is ill-defined because $X^0$
is an operator so we define the integral representation
\be
  \Theta_{\epsilon} (X^0) = -i\intR \frac{dq}{q - i\epsilon} e^{iq X^0}\quad,
\qquad \epsilon \rightarrow 0^+
\label{theta}
\ee
and then
\be
D_{\epsilon} = -i\intR \frac{dq}{q - i\epsilon}X^0 e^{iq X^0}
= -\intR \frac{dq}{(q - i\epsilon)^2} e^{iq X^0}
\ee
 where we have integrated by parts.
To find the corresponding  $C$ operator (\ref{stress}) we   study the
OPE of $D_\epsilon$ with the stress-energy tensor. 
Using   the fact that the
conformal dimension of the operator $e^{iq X^0}$ is $q
^2/2$  one  easily gets
\bea
T(w)D_{\epsilon}(z)&=& - \intR \frac{dq}{(q - i\epsilon)^2} 
~\frac{q^2}{2(w-z)^2} e^{iq X^0}   \nn \\
&=&  - \frac{1}{(w-z)^2}\intR \frac{dq}{(q - i\epsilon)^2} 
\left[(q - i\epsilon)^2 + 2i\epsilon(q - i\epsilon) -
\epsilon^2\right]e^{iq X^0}
\label{opetd}
\eea
In the first term there is no pole  and after integration we
get zero (formally it is a delta-function of $X^0$), and 
the other two  terms
give $1/q$ and $1/q^2$ poles respectively so that 
\be
T(w)D_{\epsilon}(z)= -~\frac{\epsilon^2/2}{(w-z)^2} D_{\epsilon}
+ \frac{1}{(w-z)^2}\epsilon\Theta(X^0)
\label{opetd1}
\ee
Comparing (\ref{opetd1}) with (\ref{stress}) we identify the $C$ operator 
to be  $C_{\epsilon} =
 \epsilon  \Theta_{\epsilon} (X^0)$ 
(note the factor of $\epsilon$ to which we will return).
The action of the stress tensor on the operator $C_{\epsilon}$  is:
\bea
T(w)C_{\epsilon}(z) &=& 
 -i\epsilon \intR \frac{dq}{(q -i\epsilon)} \frac{q^2/2}
{(w-z)^2)}e^{iq X^0(z)}
 \nn \\
 &=& -i\epsilon \frac{1}{2(w-z)^2}\intR 
\frac{dq}{q - i\epsilon} 
\left[(q - i\epsilon)^2 + 2i\epsilon(q - i\epsilon) -
\epsilon^2\right]e^{iq X^0}\nn\\
&=& -~\frac{\epsilon^2/2}{(w-z)^2} C_{\epsilon}
\label{opetc}
\eea
as expected.  These results show that the degenerate operators 
$C_\epsilon$ and $D_\epsilon$ have conformal dimension 
$\Delta=-{\epsilon^2\over 2}$ which is negative and 
vanishes in the
limit $\epsilon \rightarrow 0^+$ (note that this implies that the
total dimension  of the impulse operator (\ref{wrongrecoil}),
including the $\partial_{n}X^i$ factor, is $1-{\epsilon^2\over 2}$). 
 We see  that
for non-zero $\epsilon$ the impulse  operator is  relevant
in a renormalization-group sense. This is related
to the very nature of the logarithmic operators
which  lie on the border line between conformal
field theories and general two-dimensional
field theories \cite{kogmav}- \cite{ckt}, \cite{gkf}. 
It is the existence of such
relevant deformations in the recoil problem that lead
to a change of state of the $0$-brane background~\cite{kogmav,emn}.

It is clear from (\ref{opetd1}) that we cannot work just with the 
$D_{\epsilon}(X^0)$ operator because $C_{\epsilon}(X^0)$ will necessarily
 be induced by a scale transformation.
  Thus, the proper recoil operator is described
by
\bea
{\cal V}_{rec} =
\int d\tau ~ [y_i C_{\epsilon}(X^0) \partial _nX^i + 
 u_iD_{\epsilon}(X^0)
\partial _n X^i ] 
\label{recoil}
\eea
 where the coupling constants $y_i$ and $u_i$ in principle depend on
scale. As we shall show in this article this scale
dependence
may be interpreted as providing a
`target time' dependent shift
in the initial collective
coordinates of the $0$-brane, provided
one identifies the
world-sheet scale with an evolution parameter
in target space. 

We will now  derive explicit
expressions for the one and two point functions
of the operators  $D_{\epsilon}(X^0)$  and 
  $C_{\epsilon}(X^0)$  appearing in (\ref{recoil})
  and verify that these operators
have the correct properties to describe recoil.
 For calculational convenience
we will analytically continue the target time $X^0$
to a Euclidean-signature field, and only at the end
 return to the Minkowskian signature.
The  two-point function 
on the world-sheet disc is given by
\be
     {\cal G}_z \equiv  <X^0 (z) X^0(0)> \sim -2\eta \log|z/L|^2
\label{prop1}
\ee
where  $L$ is the size of the disc, and $z$ parametrizes
the world-sheet coordinates \footnote{If we map the disc
on the upper half-plane, then $z$ becomes a real variable $\tau$ in
the expression
(\ref{prop1}). This will be understood in what follows.}.
The value of $\eta $ depends on the (flat) target space signature
of $X^0$, being $+1$ for  Euclidean and $-1$ for Minkowski signature.
The coincidence limit $z \rightarrow 0$ of (\ref{prop1})
is regularized by an ultraviolet cut-off $a$ so  that :
\be
 {\cal G}_0 \equiv <X^0 (z) X^0(z)> \sim 2\eta \alpha
\qquad \alpha \equiv \log|L/a|^2
\label{coinc}
\ee
To be rigorous one should consider the covariant
propagator on the disc, of size $L$,
and then take the coincidence
limit.
However, the simplified expression (\ref{coinc})
is sufficient to yield the correct leading
behaviour in the OPE.
 
Using (\ref{coinc})
 the one point function of the operator $C_\epsilon$
is given by 
\bea <C_\epsilon> &=& -i\epsilon \intR\frac{dq}{q-i\epsilon} <e^{iqX^0}>
\nn \\
&=&\epsilon^2 \intR \frac{dq}{q^2 + \epsilon ^2}
e^{-\eta q^2\alpha }
\label{c1}
\eea
The result of the integration in (\ref{c1})
may be expressed~\cite{tables} in terms of the Error
function $\erf(x) \equiv
\frac{2}{\sqrt{\pi}} \int _0^x e^{-t^2} dt$ as:
\be
I=\int _0^\infty dx \frac{e^{-\alpha x^2}}{x^2 + \epsilon ^2}=
\frac{\pi}{2\epsilon} e^{\alpha \epsilon^2}
\left(1-\erf(\epsilon\sqrt{\alpha})\right)
\quad, \qquad {\rm Re}\,\alpha > 0
\label{integral}
\ee
Then,
\be
<C_\epsilon>=\epsilon \pi e^{\eta \epsilon ^2\alpha}
\left(1-\erf(\epsilon\sqrt{\eta\alpha})\right)
\label{c2}
\ee
We note, at this stage, that if we take the limit $\epsilon
\rightarrow 0$  so that
\be
   \epsilon^2\alpha= \epsilon ^2 \log|L/a|^2 = {\rm finite ~ constant}
\label{finite}
\ee
then $<C_\epsilon>$    (\ref{c1})
is zero  for $\epsilon \rightarrow 0$.
Similarly, 
the one point function for  $D_\epsilon$ (\ref{recoil})
is given by
\bea <\D>&=&\intR \frac{dq}{(q-i\epsilon)^2}<e^{iqX^0}> \nn\\
&=& -i\frac{\partial}{\partial \epsilon} \intR  \frac{dq }
{q - i\epsilon}
e^{-\eta q^2 \alpha} \nn \\
 &=&\frac{2}{\epsilon} \left\{ -\pi \alpha \epsilon ^2 e^{\epsilon^2 \alpha}
\left(1 -\erf(\epsilon\sqrt{\alpha})\right) + 
\epsilon\sqrt{\pi}\sqrt{\alpha}\right\}
\label{dopera}
\eea
On account of (\ref{finite}), then,
the one-point function (\ref{dopera})
of the $D_\epsilon$ operator diverges as $1/\epsilon$.

The two point function for  $\C$  is given by
\bea <\C(z)\C(0)> = -\epsilon^2 \intR \intR \frac{dq~dq'}
{(q-i\epsilon)(q' - i\epsilon)}<e^{iqX^0(z)}e^{iq' X^0(0)}>  \nn \\
 = -\epsilon^2\intR \intR \frac{dq~dq'}{(q-i\epsilon)(q' -i\epsilon)}
\exp\left(-(q+q')^2\eta \alpha\right)
\times \exp\left(2\eta qq' \log|z/a|^2\right)
\nn\\ \label{cc}
\eea
where we have used
\be  <e^{iqX^0(z)}e^{iq' X^0(0)}>=
e^{-\frac{q^2}{2}<X^0(z)X(^0z)>-
\frac{q'^2}{2}<X^0(0)X^0(0)>-
qq'<X^0(z)X^0(0)>}
\ee
and  (\ref{prop1},{\ref{coinc}).
We now observe that in the limit $\alpha \rightarrow \infty$,
with the condition (\ref{finite}) being assumed to  hold,
the first exponential would give a `smeared' $\delta (q+q')$
contribution to the integral multiplied by
 a normalization factor $\sqrt{\pi/\alpha}$. Then
 the dominant contributions to the integral in (\ref{cc}) 
as $\alpha \rightarrow \infty$ 
come from the regions in $q'$ integration for which
$q'=-q$. With this in mind we get 
\bea<\C(z)\C(0)> &\sim& -\epsilon^2\sqrt{\frac{\pi}{\alpha}} 
\intR \frac{dq}{(q^2+\epsilon^2)}
e^{-2\eta q^2\log|z/a|^2} \nn\\
&=&-\epsilon^2  \pi\sqrt{\frac{\pi}{\epsilon^2\alpha}} 
e^{2\eta\epsilon ^2 \log|z/a|^2} \left(1-\erf\left(\epsilon 
\sqrt{2\eta \log|z/a|^2}\right)\right)\nn\\
&\stackrel{\epsilon\to 0}{\sim}& 0+O(\epsilon^2)
\label{cc3}
\eea
In a similar manner one may compute the $<\C(z)\D(0)>$ 
function
\newcommand{\ggg}{2\eta\epsilon^2 \log|z/a|^2}
\bea <\C(z)\D(0)> &\sim& -\frac{\epsilon}{2}\sqrt{\frac{\pi}{\alpha}}
\frac{\partial}{\partial \epsilon} \intR \frac{dq}{q^2 + \epsilon ^2}
e^{-2\eta q^2 \log|z/a|^2}\nn \\
&=&  {\pi\over2}\sqrt{{\pi\over\epsilon^2\alpha}}
\bigg\{e^{\ggg}\left(1-4\eta\epsilon^2 \log|z/a|^2\right)
\left(1-\erf\left(\sqrt{\ggg}\right)\right)\nn\\
&\,&\qquad\qquad\qquad+2\sqrt{{\ggg\over\pi}}\,\bigg\}\nn\\
&\stackrel{\epsilon\to 0}{\sim}& {\pi\over2}\sqrt{{\pi\over\epsilon^2\alpha}}
\left(1-\ggg\right)
\label{cd}
\eea
Finally the two-point function for $\D$ is given by
\bea <\D(z)\D(0)> &=&\frac{1}{\epsilon^2} <\C(z)\D(0)> \nn \\
&\stackrel{\epsilon\to 0}{\sim}& {\pi\over2}\sqrt{{\pi\over\epsilon^2\alpha}}
\left({1\over\epsilon^2}-2\eta\log|z/a|^2\right)
\label{dd}
\eea
Thus we see that in the limit 
\be
\epsilon \rightarrow 0, \qquad
  \epsilon ^2 \log|L/a|^2 \sim O(1)
\label{epsilonlog}
\ee
we obtain the canonical two-point  correlation functions (\ref{two})
 with one exception - the singular $1/\epsilon^2$ term in $<DD>$.
Note that the singularity structure at small $\epsilon$
 of the correlation functions is unaffected by considering instead the 
connected correlation functions, as may easily be checked numerically.

Because the exact value of the numerical constant in 
(\ref{epsilonlog}) is a free parameter we may choose it at will
 (the difference between different choices can be reabsorbed in the
redefinition of the $\log z$ term) and thus establish an unambiguous
relation between $\epsilon$, the regularization parameter in a
target-space,  and $L/a$, which is a world-sheet scale. Comparing with
(\ref{prop1}) it is most natural to put  
  ${1\over\epsilon^2} = 2 \eta \log|L/a|^2$ and then we get (up to a
 normalization factor):
\bea   <C_{\epsilon}(z) C_{\epsilon}(0) > &\sim& 0 \nn \\
   <C_{\epsilon}(z) D_{\epsilon}(0) > &\sim & 1 \nn \\
    <D_{\epsilon}(z)D_{\epsilon}(0)> &\sim& -2 \eta \log|z/L|^2
\label{CD}
\eea

Now let us make a scale transformation
\be
L  \rightarrow L' = L e^{t}
\label{fsscaling}
\ee 
which is really a finite size scaling (the only one which has 
physical sense for the open string world-sheet). Because of the
 relation between
 $\epsilon$ and $L$ this transformation will change $\epsilon$
\be
\epsilon^2  \rightarrow \epsilon'^2 =
 \frac{\epsilon^2}{1 + 4\eta \epsilon^2 t}
\label{epsilontransform}
\ee
(note that if $\epsilon$ is infinitessimally small then 
$\epsilon'$ is also infinitessimally small for any finite $t$)
and we can deduce from the scale dependence of the correlation functions
(\ref{CD}) that $ C_{\epsilon}$ and $D_{\epsilon}$ transform as:
 \bea
D_{\epsilon} &\rightarrow& D_{\epsilon'} =
 D_{\epsilon} - t C_{\epsilon} \nn \\
 C_{\epsilon} &\rightarrow& C_{\epsilon'}= C_{\epsilon}
 \eea
 From this transformation
 one can see that the coupling constants in front of 
$ C_{\epsilon}$ and $D_{\epsilon}$  in the recoil operator
(\ref{recoil}), i.e. the velocities $u_i$
and spatial collective coordinates $y_i$ of the brane, must transform like:
\be
 u_i \rightarrow u_i~~,~~y_i \rightarrow y_i + u_i t 
\label{scale2}
\ee
Thus, in the presence of recoil
a world-sheet scale transformation leads to an evolution
of the $D$-brane in target space.

It is amusing that the
 pair of logarithmic operators describing 
$D$-brane recoil,  can   be used  to give a new 
 way to look at the stringy  uncertainty principle~\cite{ven}
\be 
   \Delta X_i \sim \frac{\hbar}{\Delta P_i} + O(\alpha '_s)\Delta P_i 
\label{uncert}
\ee
where $\alpha '_s$ is the string (or $0$-brane) scale (involving the 
coupling constant of the string). 
Indeed, from (\ref{finite}) 
it becomes clear that small $\epsilon$ could be 
viewed as reflecting an uncertainty in the 
  energy of the $D$ brane. 
The r\^ole of the $D$ operator
in (\ref{recoil}), then, can be easily 
checked to be in agreement with the standard 
quantum mechanical part of the uncertainty (\ref{uncert}), 
whilst 
 the $C$ operator is associated with 
essentially stringy effects.  
 To see this one  have to take into account  that for finite  $\epsilon$
the regulated $\Theta _\epsilon \sim \Theta(X^0)e^{-\epsilon X^0}$ 
 and so $\epsilon = 1/\Delta t$, where $\Delta t$ is the resolution in
 time. The $X_i$ coordinate of the $0$-brane starts to grow  
 due to the recoil  as $u_i X^0$  and the maximal value it will 
 reach will be of order
 $u_i /\epsilon$
in the  cases when $\epsilon \rightarrow 0^+$. To  connect with the
 momentum uncertainty we have to show that $\Delta P_i \sim \hbar\epsilon/u_i$.
Recall that to measure momentum in quantum theory (see, for
example \cite{LL}) one has to scatter the quantum particle ($0$-brane
in our case) on a ``detector'' and to measure both momentum and energy
 of ``detector'' before and after  scattering. The ``detector'' may be
 a closed string state with definite energy and momentum.  If  the
measurement (recoil)  takes a  finite time $\Delta t = 1/\epsilon$ the total
energy  is uncertain and one has the following relations
\bea
p +P = p' + P' \nn \\
|e + E - e' - E'| \sim \hbar\epsilon
\eea
where small letters refer to the ``detector'' and big ones the $0$-brane.
  Primed  and unprimed letters correspond to the quantities
before and after collision. Assuming that we know exactly the momentum and
energy of the detector before and after the collision we obtain the 
uncertainties
\be
\Delta P = \Delta P', ~~~~ (\Delta E - \Delta E')\sim \hbar\epsilon
\ee
But 
\be 
\Delta E = \frac{\partial E}{\partial P}  \Delta P = v\Delta P
\ee
where $v$ is a velocity and so
\be
(v'-v)\Delta P \sim \hbar\epsilon.
\ee
 In  the  case of the recoil operator (\ref{recoil}) the change in velocity is 
$u$ and finally we get
\be
 \Delta P \sim \hbar\epsilon/u \label{dph}
\ee
 i.e. precisely the  contribution of the $D_{\epsilon}$ operator.
 However if one
increases
$ \epsilon$ (making the uncertainty in momentum and energy bigger) one
 can see that the $C$ term will give a minimal bound of order of 
 $y \epsilon$ (because of the factor of $\epsilon$ which appears in the 
definition 
of $C$).  Using (\ref{dph})  we see that the second term in
(\ref{uncert}) must be ${uy\over\hbar}\Delta P$  and note that $uy$
has the same dimension as $\alpha'_s$.
Thus we see that the recoil term yields the usual
Heisenberg part of the uncertainty $1/\epsilon$,  whereas the $C$ term is a
stringy counterpart giving an $\epsilon$ term. It is important that,
because of the C-D mixing, one can not evade the influence of the $C$ term.
  The  fact that there are  two different operators in the 
  recoil operator enables  us to reproduce both terms in
(\ref{uncert}).

 Let us discuss briefly how our results for $0$-brane
 can be generalized for a $p$-brane. In this case one has $p+1$ 
 coordinates $X^{I}~, I = 0,1, \dots, p$
 with Neuman boundary conditions and $(9-p)$
 coordinates $X^i~,i=p+1,\dots 9$  
satisfying Dirichlet boundary conditions
\be
 X^i ({\rm boundary}) = 0,~i=p+1,\dots, 9
  \qquad \partial _n X^I ({\rm boundary}) = 0,~ I = 0,1, \dots, p
\label{dbrane}
\ee
The relevant Poincar\'e symmetries  are given now by translations $P^i$
 in transverse $9-p$ directions and Lorentz rotations $M^{Ij}$ which
 include   boosts in transverse directions $M^{0j}$ as well as
rotations $M^{aj},~ a = 1,\dots p;~j = p+1,\dots, 9$ in 
transverse-longitudinal planes. Let us also note that the commutators
 of these  generators
\be
[P^i, P^j] = 0,~~~[P^i, M^{Ij}] = g^{ij} P^I, ~~~
[M^{Ii}, M^{Jj}] = g^{IJ}M^{ij} + g^{ij} M^{IJ}
\ee
 corresponds to translations $P^I$ and rotations $M^{IJ}$ inside the 
 $p$-brane world-volume as well as rotations $M^{ij}$ in  transverse
planes, i.e. all these commutators do not  change the state of the
 $p$-brane itself.
This means that the result of two consecutive non-parallel recoils or
bendings, with parameters $(y_1,u_1)$ and $(y_2,u_2)$ respectively,
does not depend upon their ordering and is determined by the new set 
of parameters $(y_1+y_2,u_1+u_2)$
  Besides recoil one may consider now the bending of the
$p$-brane  in which case instead of  $\Theta _\epsilon(X^0)$ we have
 to consider $\Theta _\epsilon(X^I)$. The most general form of the 
``bending-recoil'' operator will be 
\bea
{\cal V}_{ben-rec} =
\int d\tau ~ [y_{iI} C_{\epsilon}(X^I) \partial _nX^i + 
 u_{iI}D_{\epsilon}(X^I)
\partial _n X^i ] 
\label{bendrecoil}
\eea
 where for coupling constants $y_{iI}$ and $ u_{iI}$ we will get the
same scaling relations (\ref{scale2}).

It is curious to note that there is some  similarity between
the mechanism relating world-sheet scales to target time that
 we have described here and what occurs in the Liouville sector of 
the non-critical string theory \cite{kogan,emnl}. Although the
$D$-brane calculation is formally in a critical string model the
emergence of operators which are relevant for $\epsilon >0$
may indicate a connection with non-critical  theories.
Presumably it would be necessary to formulate the $D$-brane
problem on a curved world volume \cite{emn} in order to
establish such a connection.

As a final comment, we wish to
discuss the 
mixing between $C$ and $D$
operators
in connection with the unitarity
of the effective theory containing only
low-energy point-like degrees of freedom.
We shall concentrate on the
computation of a four-point
scattering amplitude in a generic
string theory
with logarithmic operators in its spectrum.
For simplicity
we shall concentrate in bosonic strings but one should
note that string soliton backgrounds are known to be
stable (exact) solutions of conformal field theory
only in space-time supersymmetric theories.
 However 
the bosonic computation outlined
here is indicative of the sort
of modifications to the Veneziano amplitude
(of the bosonic part of the superstring)
in a generic theory with logarithmic operators.
In the compactification of such string theories
down to a four-dimensional flat space-time  the logarithmic
operators, which arise from the underlying conformal field theory
models that describe the solitonic string background~\cite{kogmav},
 are confined to
the compact dimensions of the string, and one may
assume that from a world-sheet point of view
they lead to non-trivial `bulk' contributions
to the amplitudes.
The issue we would like to address first
is whether such compactifications
are consistent  with unitarity of the spectrum
of the string.
To be precise we shall perform the computation
of correlation functions of four lowest-lying
states in an uncompactified
four-dimensional space time in an open (super)string theory
with
six dimensions compactified (a similar phenomenon takes place in the
closed string  sector but in this case there are logarithmic operators
in both left and right sectors). 
The correlation function factorizes into an ordinary
space-time part and a conformal block over the
compact space 
\be
A_4 =<V_1 V_2 V_3 V_4 >_{M_4} <O_1 O_2 O_3 O_4 >_{K_6}
\label{ampl}
\ee
where $M_4$ and $K_6$ indicate four-dimensional
Minkowski and compactified   space respectively.
The existence of logarithmic operators
in the compact sector would generically imply
the presence of $\log |z|$ terms on the world-sheet
coming from the anomalous operator product expansion
of (degenerate) primaries.
Indeed, consider the O.P.E. between
two such operators of the full ten-dimensional theory,
which we factorize into a non-compact
piece $:e^{ikX} :$ and a compact piece $O_i$,
\be
:e^{ikX(z)}::e^{ik'X(0)}:
O_i (z) O_j (0) \sim   |z|^{ k.k'- 2 }
\frac{\log|z|}{|z|^2}
e^{i(k+ k')X}O
\sim  \partial _{k.k'-2 } (|z|^{k.k' -2} ) e^{i(k+ k')X} O
\label{ope}
\ee
where for open strings the world-sheet
variables are understood as lying on the real axis.
Integrating the right-hand side around zero it is obvious
that there exists a (non-unitary) double pole,
thereby indicating the breakdown of tree-level
unitarity in this
type of compactification (!) This is an essentially
stringy effect, with no precedent in $\sigma$-model
computations.

 From our discussion in this paper, this
loss of unitarity seems to be only apparent in
certain string backgrounds involving
$D$-brane excitations. Indeed, if the compactified
soliton backgrounds are viewed as $D$-branes,
then
in the above computation
one should include quantum $D$-brane
excitations. The latter produce themselves
a mixing between boundary logarithmic operators,
which in turn
might cancel the mixing effects due to
the logarithmic operators in the bulk.
At present this problem is not solved.

\end{document}